\documentclass[twocolumn,superscriptaddress,nofootinbib,preprintnumbers]{revtex4-2}
\usepackage[colorlinks=true,breaklinks=true]{hyperref}
\usepackage[utf8]{inputenc}
\hypersetup{urlcolor=Blue,
	    citecolor=Blue,
	    linkcolor=Blue}
\usepackage{mathtools}
\usepackage[dvipsnames]{xcolor}
\usepackage{bbm}
\usepackage{orcidlink}

\usepackage{amsmath}
\usepackage{amssymb}
\usepackage{wasysym}

\usepackage{xurl} 
\usepackage{fontawesome5} 
\DeclareDocumentCommand{\GitHub}{s}{%
	\IfBooleanTF{#1}{%
		\faGithub~\url{https://github.com/xunjiexu/MESA-DP.git}%
	}{%
		\href{https://github.com/xunjiexu/MESA-DP.git}{\faGithub}%
	}%
}

\newcommand{\lettersection}[1]{\noindent{\bf \emph{#1}}\,---\,}

\begin{document}

\title{Dark Photons from Red Dwarfs}

\author{Stefan Vogl \orcidlink{0000-0002-3005-9279}}
\email{stefan.vogl@physik.uni-freiburg.de}
\affiliation{Institute of Physics, University of Freiburg\\Hermann-Herder-Str.~3, 79104 Freiburg, Germany}

\author{Xun-Jie Xu \orcidlink{0000-0003-3181-1386}}
\email{xuxj@ihep.ac.cn}
\affiliation{Institute of High Energy Physics, Chinese Academy of Sciences\\ Beijing 100049, China}
\preprint{...}
\date{\today}

\begin{abstract}

Light dark photons can be produced in stellar systems and thus contribute to the stellar cooling rate. 
The additional cooling changes the evolution  of the star and has an impact on various observable properties such as radius, photon luminosity or the emitted neutrino fluxes. 
This has been exploited before to derive limits based on observations of the Sun, horizontal branch stars and red giants. Given the wealth of astrophysical data collected in the last decade and the improvements in modeling stellar evolution it is interesting to investigate whether other stellar systems offer a complementary avenue towards testing dark photons. 
In this work, we study the effect on an alternative class of stars. We focus on the impact of dark photon induced cooling on red dwarfs, i.e. the lowest mass stars on the Hydrogen main sequence. Running simulations of the evolution of red dwarfs with dark photon cooling we determine the impact on the mass-radius relation. 
Combining our simulations with precise determination of mass and radius derived from observations of eclipsing binaries that have recently become available,  
we derive competitive limits which outperform the solar ones in a significant part of the parameter space. 

\end{abstract}
 
\maketitle


\lettersection{Introduction} 
Dark photons are an intriguing extension of the Standard Model (SM) since they interact via one of the few dimension-4 portals 
between the SM and new particles. 
This construction has received a lot of interest since the pioneering work of \cite{Holdom:1985ag,Okun:1982xi} with a flurry of activity in the last decade. 
Because dark photons interact feebly, which permits free-streaming through most media, and can convert into standard photons (and vice versa) in media, they exhibit a rich phenomenology in astrophysics~\cite{Redondo:2008aa,An:2013yfc,Redondo:2013lna,Redondo:2015iea,Giannotti:2015kwo,Vinyoles:2015aba, Hardy:2016kme,An:2020bxd,Li:2023vpv,Dolan:2023cjs,Vogl:2024ack,Chang:2016ntp,Rrapaj:2015wgs,Kazanas:2014mca,Dent:2012mx} and cosmology~\cite{Redondo:2008ec,Fradette:2014sza,Berger:2016vxi,Pospelov:2018kdh,Ibe:2019gpv,McDermott:2019lch,Coffey:2020oir,Li:2020roy,Caputo:2022keo,Adshead:2022ovo,Pirvu:2023lch,Gan:2023wnp,Cyncynates:2023zwj,Aramburo-Garcia:2024cbz,McCarthy:2024ozh,Trost:2024ciu,Cyncynates:2024yxm,Xu:2025wlq,Hook:2025pbn,Jaeckel:2008fi}.
An overview of the current status of the field and recent developments can e.g. be found in the reviews~\cite{Fabbrichesi:2020wbt,Caputo:2021eaa,Caputo:2025avc,Caputo:2026pdw}.   

At low energies the relevant part of the Lagrangian of the model is given by 
\begin{align}
\label{eq:lagrangian}
	 \mathcal{L} \supset   -\frac{\epsilon}{2} F_{\mu \nu} V^{\mu \nu} + \frac{m_{\rm DP}^2}{2} X_\mu X^\mu\,,
\end{align}
where $F_{\mu\nu}= \partial_\mu A_\nu - \partial_\nu A_\mu$ and $V_{\mu\nu}= \partial_\mu X_\nu - \partial_\nu X_\mu $ denote the field strength tensors of the photon and dark photon respectively.
The in-vacuum mass of the dark photon is given by $m_{\rm DP}$. The fields interact via the kinetic mixing term $F_{\mu \nu} V^{\mu \nu}$ with a strength that is controlled by the kinetic mixing parameter $\epsilon$.

Stringent stellar limits on the interaction strength of a light dark photon with $m_{\rm DP}\lesssim 10^4$ eV can be derived from observations of stellar systems, most notably the Sun, horizontal branch stars and red giants~\cite{Redondo:2008aa,An:2013yfc,Redondo:2013lna,Redondo:2015iea,Giannotti:2015kwo,Vinyoles:2015aba, Hardy:2016kme,An:2020bxd,Li:2023vpv,Dolan:2023cjs}. 
In this letter, we ask whether other stellar systems can be competitive in testing dark photon physics. 
We start with the following general considerations.

The luminosity of main-sequence stars $L_\ast$ is a strongly growing function of the stellar mass $M_\ast$. 
This can be easily understood since the condition of hydrostatic equilibrium requires that the increasing gravitational force is countered by an increase in pressure, which is accompanied by an increase in energy production. 
For a simplified star modeled as an ideal gas with constant opacity,  the mass-luminosity relation is  given by  $L_\ast \propto M_\ast^\eta$ with $\eta=3$ \cite{Kippenhahn_Weigert_Book}, whereas for realistic stars $\eta$ has some mass dependence. 

In contrast, the luminosity of a star in exotic light particles can have a very different dependence on the stellar mass.  In the low mass limit characterized by $m_{\rm DP}\ll \omega_p$ where  $\omega_p$ is the plasma frequency in the stellar interior, the emission of dark photons is dominated by the longitudinal polarization of the dark photon \cite{An:2013yfc}. The longitudinal mode can be produced resonantly with $\omega\approx \omega_p$ 
if $m_{\rm DP}\leq \omega_p$. 
Under these conditions, the dark photon luminosity follows~\cite{An:2013yfc,Redondo:2013lna}
\begin{align}
L_{\rm DP}\propto \int_0^{r_{\rm max}} dr r^2 \frac{\omega_p^3}{e^{\omega_p/T}-1}\,,
\end{align}
where both $\omega_p$ and the temperature $T$ are implicit functions of the radius $r$ and $r_{\rm max}$ is the largest radius for which $m_{\rm DP}\leq \omega_p$. 
Noting that $\omega_p^2=4 \pi \alpha n_e/m_e$ 
is proportional to the electron number density $n_e$ which is in turn proportional to the density 
$\rho$, we can estimate that the luminosity-mass relation for dark photons will roughly follow 
$L_{\rm{DP}}\propto M_\ast^{\eta_{\rm DP}} $ with
$\eta_{\rm{DP}}\sim 1 -1.5$.
This is a much weaker growth than for photons and, consequently, we are led to expect that on the main sequence the relative dark photon luminosity $L_{\rm{{DP}}}/L_\ast$ is maximized for the lightest stars (i.e., red dwarfs). 
We illustrate this argument in the upper panel of Fig.~\ref{fig:mesa-result} (see also the left panel of Fig.~\ref{fig:Sun} in the supplemental material for an individual breakdown to $L_\ast$ and $L_{\rm{DP}}$). 
Although the actual luminosity-mass relation is more complicated than a simple power law 
it confirms the overall expectation that $L_{\rm DP}/L_\ast$ is maximized for low-mass stars. 
As can be seen, the fractional luminosity $L_{\rm DP}/L_\ast$ in stars with $M_{\ast} \lesssim 0.3 M_\odot$ is more than one order of magnitude higher than that of a solar-mass star.
Therefore, we focus on this mass range here and study stars with masses from $0.1M_\odot$ to $0.3M_\odot$. This mass range also provides additional benefits since it combines a relatively simple stellar evolution with good observational data. For lower masses, data becomes scarce and less precise, while stars with $M_{\ast} \gtrsim0.35 M_\odot$ transition from a fully-convective interior to a mixed structure with both radiative and convective zones, which makes tracking their evolution more involved and adds additional modeling uncertainties.

While the enhanced relative new physics luminosity of low-mass stars for some classes of new physics has been noted in the past (see e.g.~\cite{Frieman:1987ui}), this idea has not been pursued before to the best of our knowledge. The reasons for dismissing low-mass stars as test systems for light new physics were the scarcity and imprecision of astrophysical data and the difficulties in modeling their stellar evolution. We want to argue that both of these issues have been overcome by now. With both high-quality data and accurate stellar models, red dwarfs provide a promising target for new physics searches.

\begin{figure}
\centering
\includegraphics[width=0.49\textwidth]{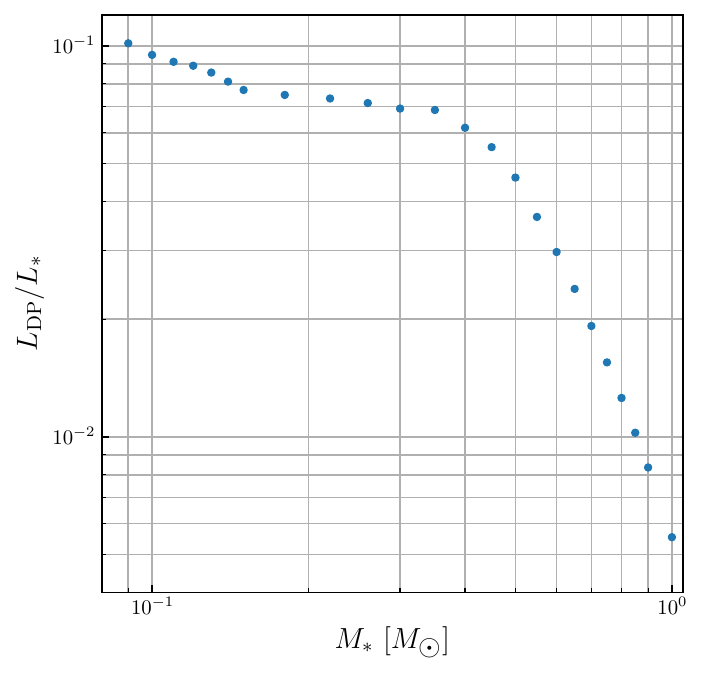}
\includegraphics[width=0.49\textwidth]{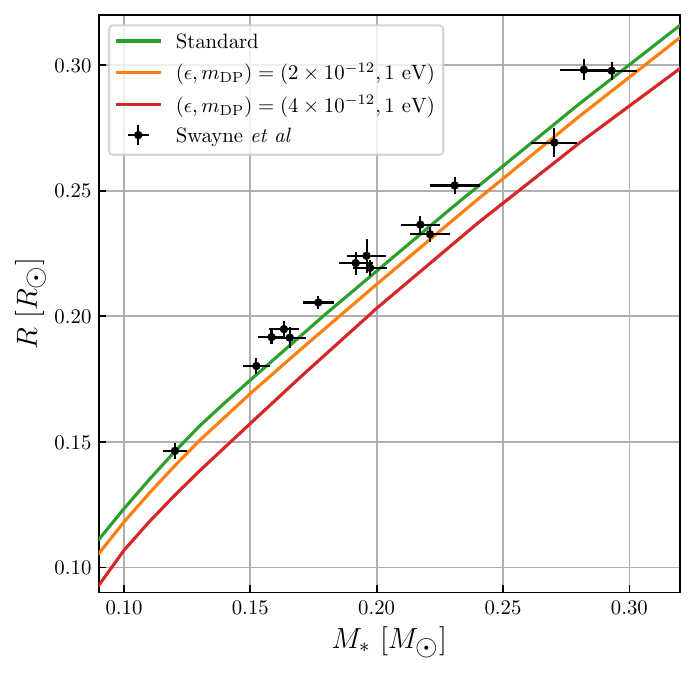}
\caption{Upper Panel: MESA results of the dark photon luminosity $L_{\rm DP}$ divided by the standard stellar photon luminosity $L_{*}$ for a representative choice of $m_{\rm DP}=1 {\rm eV}$ and $\epsilon =1\times 10^{-12}$.  Note that $\epsilon$ has been chosen such that the overall structure of the stars is not perturbed by the additional cooling.
Lower Panel: The mass-radius relation obtained from MESA simulation compared with the red dwarf data from Swayne {\it et al}~\cite{2024MNRAS.528.5703S}. 
\label{fig:mesa-result}}
\end{figure}

\lettersection{Dark photon cooling rates}
The presence of the kinetic mixing in Eq.~\eqref{eq:lagrangian} leads to non-standard kinetic terms for both the dark and the regular photon. This can be resolved by an appropriate field redefinition.
Note that in the presence of a dense standard model plasma, the propagation eigenstates depend on the medium and kinetic theory or thermal field theory has to be used to derive the production rates of the dark photon, see \cite{Redondo:2013lna} for a comparison of the two approaches.
In the interior of stars, 
the crucial quantity that controls these effects is the plasma frequency $\omega_p$.
The emission rate of the dark photons depends sensitively on their polarization.  Longitudinal modes are strongly preferred for $m_{\rm DP}\ll \omega_p$ since their production rate is proportional to $m_{\rm DP}^2$ \cite{An:2013yfc} while the production of 
the transverse mode in the low-mass limit is proportional to $m_{\rm DP}^4$ \cite{Redondo:2008aa}. 
Both the longitudinal and the transverse dark photon emission rates are strongly dominated by resonant phenomena albeit with important quantitative and qualitative differences. In the following, we briefly sketch the resonant contributions, 
and refer to Refs.~\cite{Redondo:2008aa,An:2013yfc,Redondo:2013lna} for detailed discussions on the emission rate including non-resonance contributions---see also \ref{appx:rates} for a brief review of the formalism.

In the case of the longitudinal mode with $m_{\rm DP} \leq \omega_p $, the emission of dark photons with $\omega \sim \omega_p$ is strongly enhanced. In this case,  the energy loss rate per unit volume is well approximated by \cite{An:2013yfc,Redondo:2008aa}
\begin{align}
    Q_{\rm L}= \frac{\epsilon^2 m_{\rm DP}^2}{e^{\omega_p/T}-1} \frac{\omega_p^3}{4 \pi} \sqrt{1-m_{\rm DP}^2/\omega_p^2} \, .
\end{align}
As $\omega_p\propto \sqrt{n_e}$ and the density is a monotonically decreasing function of the radius, 
the resonant production in the longitudinal mode is allowed up to a maximal radius 
$r_{\rm max}$ determined by $\omega_p(r_{\rm max})=m_{\rm DP}$. 
Therefore, this is a spread-out volume emission.

The transverse mode can also be produced resonantly but both the structure of the coupling and the properties of the resonance are different; see  \cite{Redondo:2008aa,Redondo:2015iea} for an in-depth discussion of the transverse mode. For $m_{\rm{DP}}=\omega_p$ every momentum mode is produced resonantly simultaneously. However, since this condition is only met in one mass shell, if at all, the emission is concentrated in a small volume element. Keeping only the resonant contribution and using a narrow width approximation one finds \cite{Redondo:2008aa} \footnote{This result differs from the equivalent approximation given in Eq.~(27) of \cite{Redondo:2008aa}  by a factor $\pi/2$. This can be traced to an imprecise handling of the narrow width approximation in \cite{Redondo:2008aa}. }
\begin{align}
    Q_{\rm T}= 
    \frac{\epsilon^2 m_{\rm DP}^4}{ \pi} \int d\omega \frac{\omega \sqrt{\omega^2-m_{\rm DP}^2}}{e^{\omega/T}-1} \frac{\delta(r-r_{\rm res}) }{|d \omega_p^2/dr|} \,,
\end{align}
where it is understood that the trivial integration over $r$ still needs to be performed.
This energy loss rate is of higher power in $m_{\rm DP}$ and, therefore, the longitudinal mode dominates for small $m_{\rm DP}$. Nevertheless, the transverse mode can become important and even dominant when $m_{\rm DP}$ approaches $\omega_{p,{\rm max}}$, i.e. for masses of the order of the highest plasma frequency encountered in the star. We note here that $\omega_{p,{\rm max}}$ generally increases as the stellar mass decreases---see Fig.~8 in Ref.~\cite{Nguyen:2023czp}. Consequently, red dwarfs can resonantly produce dark photons slightly heavier than those attainable in the Sun.

\lettersection{Stellar structure and the effect of additional cooling} 
We simulate the stellar evolution of red dwarfs with 
the open source stellar evolution code MESA \cite{MESA_paper,MESA_web}. 
This tool is very flexible and allows the inclusion of user-defined cooling rates.  Previous studies of additional cooling with MESA simulations were presented in \cite{Friedland:2012hj,Dolan:2021rya} for axions  and \cite{Dolan:2023cjs} for dark photons with the code related to the cooling implementation employed by these publications available at \cite{Friedland_web,MESA_ALPs,MESA_DP}. These implementation are based on old versions of MESA and do not work off-the-shelf with the current version. Therefore, we used our own implementation of the cooling rates, and made the code publicly available \GitHub.

We focus on light main-sequence stars with $0.1M_\odot <M_{\ast}< 0.30 M_\odot$, which has not been considered in previous studies. Hence, we believe that this work presents the first stellar evolution studies of low-mass stars with additional cooling.
These stars provide a good target since they are light compared to the Sun, which maximizes the relative importance of the cooling rate, and have a relatively simple evolution. They are fully convective, i.e., energy transport is dominated by convection processes throughout. This implies that they are well mixed such that their chemical composition does not vary significantly as a function of radius. Thus, the metallicity of the photosphere, which is relatively easy to  determine  observationally, is representative for the whole interior. This removes one of the key uncertainties in the modeling of solar-mass stars like the Sun. In addition, the lifetime of such stars exceeds the age of the universe significantly, which implies that their properties remain very stable once they have reached the main sequence. Thus, their precise stellar age is not a relevant parameter here and stars with ages of $1-10$ Gyr can be treated on an equal footing. 
We validate the results of our MESA runs by comparing the mass-radius relation without dark photon cooling with independent evolutionary models published in \cite{Baraffe_15}. The agreement is excellent in the mass range of interest to us. 

The mass-radius relation at solar metallicity without extra cooling, together with two representative choices of the dark photon parameters, is shown in the lower panel of Fig.~\ref{fig:mesa-result}.
As can be seen, dark photon cooling leads to a contraction of the star. This is expected and can be understood from the need to achieve a higher core temperature in order to partially compensate the   energy loss via dark photons while maintaining a sufficient photon luminosity for hydrostatic equilibrium. 
For comparison, the masses and radii of 15 stars measured by \cite{2024MNRAS.528.5703S} are also shown; see the next section for a detailed discussion of the data.

\lettersection{Astrophysical data}
As discussed in the previous section, models of stellar evolution reveal tight correlations between the fundamental properties of the star, i.e., its mass and composition, and its observable properties such as its radius and luminosity. One challenge in exploiting these connections is the need to obtain high-quality, independent measurements of these quantities. While the luminosity and the metallicity can be obtained from standard astrophysical observations, measuring radius and mass directly is more difficult. In particular, the mass is challenging since it cannot by measured directly for an isolated star and has to be inferred from its emission with varying degrees of input from stellar evolution models. An independent mass measurement thus requires a bound system that allows the determination of the stars mass from its kinematics \cite{Binney_Galactic_Astronomy}.  
The gold standard in the field
are eclipsing binaries which allow simultaneous independent determinations of both mass and radius with high accuracy, see e.g.~\cite{Serenelli_2021} for a recent overview of various methods of mass determination. There has been significant interest in scrutinizing the properties of low-mass stars recently since these are frequently hosts to exoplanets.

We use results for the mass and radius of red dwarfs published in \cite{2024MNRAS.528.5703S}. The fundamental parameters were derived from CHEOPS observations of eclipsing binaries and the data was expressly analyzed to scrutinize that mass radius relation of low mass stars. This includes the modeling of star spots, which have long been suspected to affect extraction of $M_{\ast}$ and $R$ from the light curves of eclipsing binaries. 
As we want to compare the observations with results for the stellar evolution of isolated stars, we exclude stars with a short orbital period of $P< 5$ days. 
Such stars are expected to be influenced by their more massive companion since such systems are usually tidally locked which leads to high rotation speeds of the lighter companion we are interested in. 
In addition, the radial separation in such low orbital period systems is  small such that the irradiation from the heavier companion becomes significant which potentially affects the standard cooling rate. Restricting our sample to stars with $0.1 M_\odot \leq M_{\ast}\leq 0.3 M_\odot$ and $P\geq 5$~days leaves us with 15 test stars.  The errors on $M_{\ast}$ and $R$ are at the percent level.
Therefore, we expect that this data set is sensitive to cooling-induced variations of the mass radius relation at that level.
A glance at the lower panel of Fig.~\ref{fig:mesa-result} reveals that this star sample is quite consistent with the standard mass-radius relation and there is no obvious preference for dark photon cooling. For convenience, we have also collected the relevant parameters of the stars shown here in Tab.~\ref{tab:swayne} in the supplemental material.

\lettersection{Limits}
In order to set a limit on our new physics parameters, we perform a log-likelihood ratio analysis, see e.g.~the Statistics section of \cite{ParticleDataGroup:2024cfk} for a detailed discussion. We assume independent gaussian probability distributions for the mass and the radius such that the likelihood fulfills

\begin{align}
-2\ln L(\theta)= \sum_{i}\left[\frac{R_{i}-R(\theta,m_i)}{\sigma_{R,i}}\right]^{2}+\left(\frac{M_{i}-m_i}{\sigma_{M,i}}\right)^{2}
\end{align}
where $R_i$ and $M_i$ are the observed values of the radius and mass of star $i$, with their uncertainties given by  $\sigma_{R_i}$ and $\sigma_{M_i}$, respectively. The true stellar mass $m_i$ is an input in MESA simulations while the theoretical value $R$ is a function of 
new physics parameters $\theta=(m_{\rm DP},\epsilon)$ and stellar input parameters such as the mass and metallicity. As the $m_i$ are not known we treat them as nuisance parameters in our analysis \footnote{Within the allowed errors, metallicity only has a mild impact here and one can set it to the best fit value without affecting the results.}.
 Following Wilks theorem, we assume that the log-likelihood ratio $\lambda=-2\log (L(\hat{\theta})/L(\theta)$), where $\hat{\theta}$ denotes the set of parameters that maximizes the likelihood, follows a $\chi^2$ distribution with number of degrees of freedom 
 equal to the number of model parameters  \cite{Cowan:2010js}.

\begin{figure}
\centering
\includegraphics[width=0.49\textwidth]{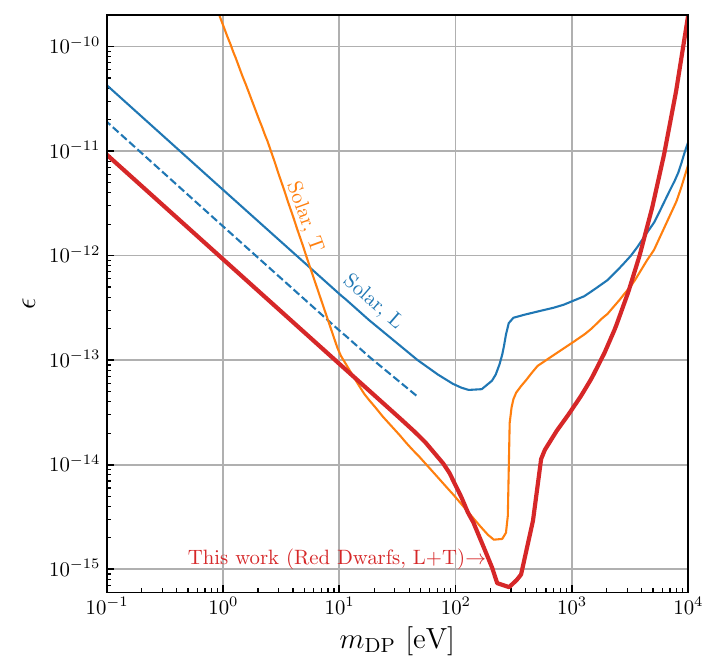}
\caption{Constrains on the dark photon from red dwarfs (red, derived in this work) and the Sun (blue and orange lines, derived in Ref.~\cite{Redondo:2013lna} assuming $L_{\rm DP}=0.1 L_{\astrosun}$). The dashed blue line corresponds to $L_{\rm DP}=0.02 L_{\astrosun}$ as derived by \cite{Vinyoles:2015aba} for the low mass limit---see the main text for further details.
\label{fig:final-result}}
\end{figure}

Our dark photon limit derived from the red dwarf data set is presented as the red curve in Fig.~\ref{fig:final-result}.
In the low-mass regime, our analysis sets an upper limit on the effective coupling $\epsilon \times m_{\rm{DP}} \leq 0.9 \times 10^{-12} $eV for $m_{\rm DP}\lesssim 10$ eV. 
At  higher masses, resonant production of the transverse mode becomes important and the limits develops a non-trivial mass dependence. 

For comparison we also include previous limits from 
solar observations.
In Ref.~\cite{Redondo:2013lna}, $L_{\rm{DP}}\leq0.1 L_\odot$ was used to set solar limits on dark photons  corresponding to the solid blue and orange curves in Fig.~\ref{fig:final-result}.  This upper limit on $L_{\rm{DP}}$ is based on generalizing results from self-consistent simulations of axion-induced cooling \cite{Schlattl:1998fz} as discussed in~\cite{Gondolo:2008dd}. 
In Ref.~\cite{Vinyoles:2015aba}, an updated analysis has been performed, using the GARSTEC stellar evolution code \cite{GARSTEC} for simulation,  and confronting  it with neutrino and helioseismological observations. They derived a limit of $L_{\rm{DP}}\leq0.02 L_\odot$ and $\epsilon \times m_{\rm{DP}} \leq 1.8 \times 10^{-12} $eV. Note that this limit only applies to the longitudinal mode in the low-mass regime. One could speculate that rescaling the limits of \cite{Redondo:2013lna} to $L_{\rm{DP}}\leq0.02 L_\odot$ will cover the main effect. However, such a procedure would not be justified for the transverse mode since here the resonance structure singles out a preferred radius for the bulk of the cooling. As the impact of the transverse mode on the observables used by \cite{Vinyoles:2015aba} remains unclear, we only present this limit  in the low-mass regime by the blue dashed line.

As can be seen, our bound outperforms previous solar limits at low masses ($m_{\rm{DP}}\lesssim 10$ eV) and in a high mass window from a few$\times 100$ eV to a few$\times 1000$ eV. 
This second window arises due to the higher density in the red dwarf core(s) which lead to slightly higher $\omega_{p,\max}$ than in the Sun.

\lettersection{Summary} 
In this work we studied the implications of light dark photons for the evolution of low mass stars on the main sequence, the so-called red dwarfs. Noting that the photon luminosity scales more steeply with the mass of the star than the dark photon luminosity, we are led to expect that the relative importance of dark photon induced cooling is more pronounced in lighter stars. The additional energy loss channel changes the evolution of these stars and affects their structure. This feeds into observable properties such as their luminosities or radii. Simulating fully convective stars in the mass range $0.1 M_\odot$ to $0.3 M_\odot$ with 
the stellar evolution code MESA, which allows the inclusion of user-defined cooling rates, 
we determine the cooling effects on the mass-radius relation. We find that the emission of dark photons leads to a reduction of the radius as expected. By comparing our results with observations of a representative set of 15 low-mass stars we are able to establish competitive limit on the dark photon parameters. In the low-mass regime, our upper limit on the effective coupling $\epsilon \times m_{\rm{DP}} \leq 0.9 \times 10^{-12}$ eV outperforms previous limits from the Sun, as shown in Fig.~\ref{fig:final-result}. 
Currently, there is considerable interest in the fundamental parameters of red dwarfs due to their prominent role as hosts of exoplanets. Therefore, we expect that detailed studies of additional systems will become available in the near future, potentially strengthening these limits further.
The implications of the red dwarf mass-radius relation for other new physics candidates, such as the axion, ALPs or a light $Z'$ will be investigated more systematically in a forthcoming publication.

{\bf Acknowledgement}
The work of S.\,V. has  been partially supported by the German Research Foundation (DFG)
via the Individual Research Grant 496940663. The work of X.\,J.\,X. is supported in part by the National Natural Science Foundation of China (NSFC) under grant No.~12141501 and also by the CAS Project for Young Scientists in Basic Research (YSBR-099). 

\bibliography{refs.bib}
\bibliographystyle{utphys}

\clearpage
\newpage
\onecolumngrid


\setcounter{equation}{0}
\setcounter{figure}{0}
\setcounter{table}{0}
\setcounter{page}{1}
\setcounter{section}{0}

\makeatletter
\renewcommand{\thesection}{S\arabic{section}}
\renewcommand{\theequation}{S\arabic{equation}}
\renewcommand{\thefigure}{S\arabic{figure}}
\renewcommand{\thetable}{S\arabic{table}}

\renewcommand{\theHfigure}{S\arabic{figure}}
\renewcommand{\theHtable}{S\arabic{table}}
\renewcommand{\theHequation}{S\arabic{equation}}
\makeatother

\begin{center}
    \textbf{\large Supplemental Material: Dark Photons from Red Dwarfs} \\ 
    \vspace{0.5cm}
    {Stefan Vogl, Xun-Jie Xu}
\end{center}

\section{MESA implementation details}
\label{appx:MESA}
\subsection{Formulae and approximations}
\label{appx:rates}
 The differential dark photon luminosity in stellar medium is given
by
\begin{equation}
\frac{dL_{{\rm DP}}}{drdk}=4\pi r^{2}\frac{k^{2}\omega}{2\pi^{2}}\Gamma_{\rm DP}^{{\rm gain}}\thinspace,\label{eq:-31}
\end{equation}
with
\begin{equation}
\Gamma_{\rm DP}^{{\rm gain}}=f_{\gamma}\Gamma_{\gamma}\frac{\varepsilon^{2}m_{\rm DP}^{4}}{\left(m_{\rm DP}^{2}-\text{Re}\Pi_{\gamma\gamma}\right)^{2}+\omega^{2}\Gamma_{\gamma}^{2}}\thinspace,\label{eq:-32}
\end{equation}
and

\begin{equation}
\text{Re}\Pi_{\gamma\gamma}=\omega_{p}^{2}\eta_{\rm LT}\thinspace,\ \ \eta_{\rm LT}\equiv\begin{cases}
\frac{\omega^{2}-k^{2}}{\omega^{2}} & \text{(L mode)}\\
1 & \text{(T mode)}
\end{cases}\thinspace.\label{eq:-33}
\end{equation}
Here, $f_{\gamma}=1/\left(e^{\omega/T}-1\right)$, $\omega_{p}^{2}=4\pi\alpha n_{e}/m_{e}$,
and $\Gamma_{\gamma}\equiv\Gamma_{\gamma}^{{\rm loss}}-\Gamma_{\gamma}^{{\rm gain}}$
with $\Gamma_{\gamma}^{{\rm loss}}$ and $\Gamma_{\gamma}^{{\rm gain}}$
the loss and gain rates of $\gamma$. In thermal equilibrium, they
are related by $\Gamma_{\gamma}^{{\rm gain}}=f_{\gamma}\Gamma_{\gamma}$
and $\Gamma_{\gamma}^{{\rm loss}}=(1+f_{\gamma})\Gamma_{\gamma}$.
In practice, $\Gamma_{\gamma}$ is often determined from $\Gamma_{\gamma}^{{\rm loss}}$,
which has been calculated in Refs.~\cite{Redondo:2008aa,Redondo:2013lna}.

Although the above formulae allow for a straightforward calculation
of the dark photon luminosity, it is important to notice that Eq.~\eqref{eq:-32}
contains a resonance when $m_{\rm DP}^{2}$ approaches $\text{Re}\Pi_{\gamma\gamma}$.
This resonance can be very narrow, rendering  numerical integrations
of $k$ or $r$ computationally expensive. It is therefore often a practical
treatment to assume that the narrow resonance can be approximated
by a Dirac delta function: 
\begin{equation}
\Gamma_{\rm DP}^{{\rm gain}}\approx f_{\gamma}\varepsilon^{2}m_{\rm DP}^{4}\frac{\pi}{\omega}\delta\left(m_{\rm DP}^{2}-\text{Re}\Pi_{\gamma\gamma}\right).\label{eq:-34}
\end{equation}
Under this approximation, $\Gamma_{\rm DP}^{{\rm gain}}$ becomes
independent of $\Gamma_{\gamma}$. 

Substituting Eq.~\eqref{eq:-34} into Eq.~\eqref{eq:-31} and integrating
over $k$, one obtains 
\begin{align}
\frac{dL_{{\rm DP}}^{{\rm (L)}}}{dr} & \approx\varepsilon^{2}m_{{\rm DP}}^{2}r^{2}f_{\gamma}(\omega_{p})\omega_{p}^{2}k_{\text{res}}\thinspace,\label{eq:-35}\\
\frac{dL_{{\rm DP}}^{{\rm (T)}}}{dr} & \approx g^{{\rm (T)}}2\varepsilon^{2}m_{{\rm DP}}^{4}r_{\text{res}}^{2}\frac{\delta\left(r-r_{\text{res}}\right)}{\left|d\omega_{p}^{2}/dr\right|_{\text{res}}}T_{\text{res}}^{3}{\cal Z}\left(\frac{m_{{\rm DP}}}{T_{{\rm res}}}\right),\label{eq:-36}
\end{align}
where the superscripts (L) and (T) indicate longitudinal and
transverse polarizations, respectively. The factor $g^{\rm (T)}=2$ accounts
for two transverse modes. The subscript ``res'' indicates that the
quantity to be evaluated at the resonance. The ${\cal Z}$ function
in Eq.~\eqref{eq:-36} is defined as
\begin{equation}
{\cal Z}\left(x\right)\equiv\int_{0}^{\infty}\frac{\kappa^{2}}{\exp\left(\sqrt{\kappa^{2}+x^{2}}\right)-1}d\kappa\thinspace,\label{eq:-38}
\end{equation}
and can be computed numerically.  It has the asymptotical behavior
${\cal Z}(x\to0)=2\zeta(3)\approx2.404$ and ${\cal Z}(x\gg1)\approx0$. 

The L-mode resonance is reached when $k$ approaches $k_{\text{res}}=\sqrt{\omega_{p}^{2}-m_{\rm DP}^{2}}$.
Thus it occurs everywhere in the stellar medium as long as  $\omega_{p}(r)>m_{\rm DP}$.
The T-mode resonance only occurs at $r=r_{\text{res}}$, where
$r_{{\rm res}}$ is obtained by solving $\omega_{p}(r)=m_{\rm DP}$. 
When $m_{\rm DP}$ is small, Eq.~\eqref{eq:-35} reduces
to Eq.~(4.1) in \cite{Redondo:2013lna}.

\begin{figure*}
\centering
\includegraphics[width=0.478\textwidth]{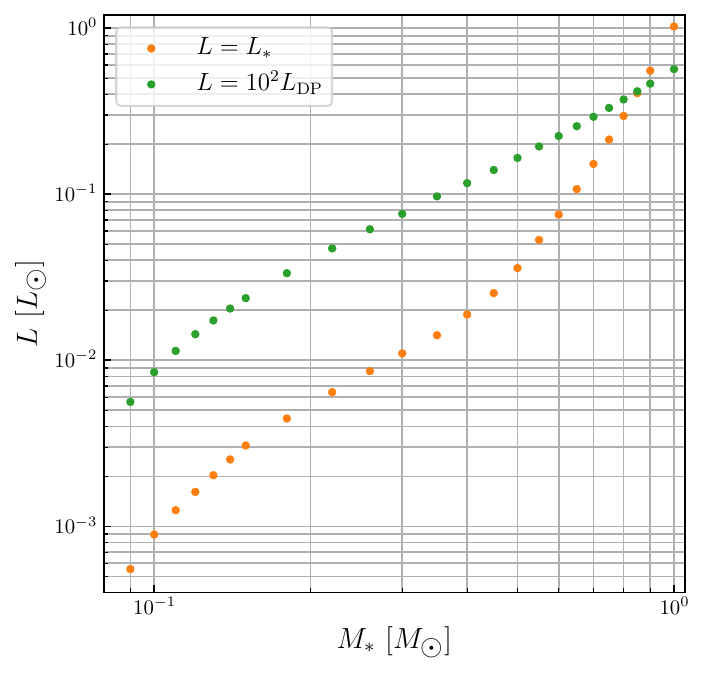}
\includegraphics[width=0.49\textwidth]{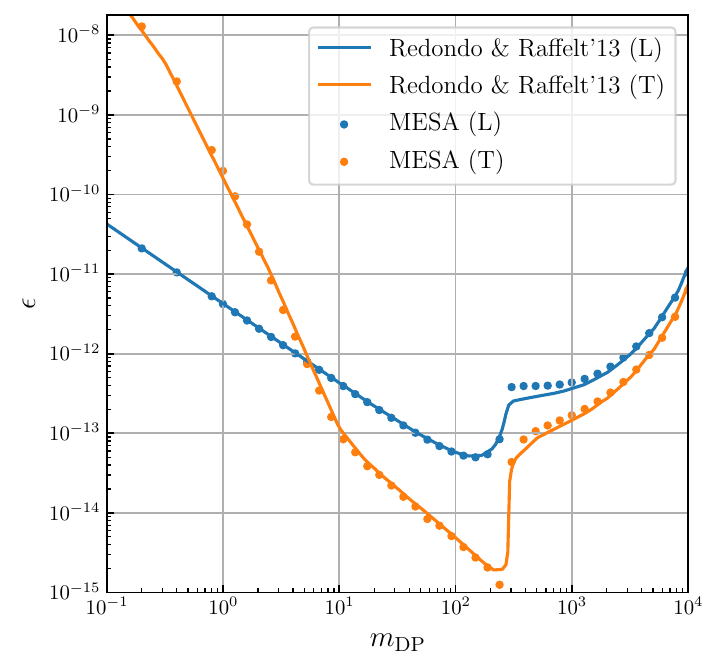}
\caption{\label{fig:Sun} Left Panel: Luminosities of photons ($L_{*}$) and dark photons ($L_{\rm DP}$) obtained from our MESA simulation of stars with $M_{\ast}\in [0.09,1.0]M_{\astrosun}$, 
assuming $(m_{\rm DP}, \epsilon)= (1~{\rm eV}, 10^{-12})$ for illustration. 
Right Panel: The $L_{\rm DP}=0.1L_{\astrosun}$ solar bound on $\epsilon$ obtained from our MESA simulation
 (blue and orange points for $L$ and $T$ modes), compared
with the earlier calculation in Ref.~\cite{Redondo:2013lna} (blue
and orange lines). }
\end{figure*}

In our MESA implementation, we first use $\omega_{p}(r)$ to infer
whether a specific stellar medium layer ($r\in[r_{i},r_{i+1}]$) contains
an L- or T-mode resonance. If a resonance is detected, we use
Eq.~\eqref{eq:-35} or \eqref{eq:-36} to calculate the local energy
loss. If it is resonance-free, we use the more straightforward approach,
i.e., Eqs.~\eqref{eq:-31} and \eqref{eq:-32}, to compute the energy
loss. In the left panel of Fig.~\ref{fig:Sun}, we present the photon and dark photon luminosities obtained from our MESA simulation. 
To verify the validity of our MESA implementation, we run the
code to compute the solar production of dark photons for L- and T-modes separately and compare the $L_{{\rm DP}}=0.1L_{\astrosun}$ contours with 
with those in Ref.~\cite{Redondo:2013lna}, as shown in the right panel of Fig.~\ref{fig:Sun}.
The solid curves represent the required coupling strength to achieve
$L_{{\rm DP}}=0.1L_{\astrosun}$, taken from Ref.~\cite{Redondo:2013lna}.
The discrete points represent our results obtained by running the
MESA code. The two results are in excellent agreement despite the fact that~\cite{Redondo:2013lna} does not include the feedback of the cooling on the stellar structure.

\subsection{Other details}

Our code is publicly available on GitHub~\GitHub. There are a few
other details worth mentioning in the MESA implementation. 

We implement the energy loss due to dark photons by first turning
on the \verb|other_energy| hook in \nolinkurl{inlist_project} and
then assigning an energy loss function to \verb|s% other_energy|
in \nolinkurl{run_star_extras.f90}, using the template 
\nolinkurl{$MESA_DIR/include/standard_run_star_extras.inc}.
Further details can be found in the tutorial~\cite{MESA_tutorial}.
Despite the suggestive name \verb|other_energy| accepts both positive and negative energy contributions, corresponding
to production and absorption of extra energy, respectively. MESA incorporates
the extra energy assigned to \verb|s% other_energy| 
\nolinkurl{sources_ad = eps_nuc_ad - non_nuc_neu_ad + extra_heat_ad + ...}
in \nolinkurl{$MESA_DIR/star/private/hydro_energy.f90}.

Note that, in principle, one could alternatively use \verb|s% other_neu| from the
template \nolinkurl{$MESA_DIR/star/other/other_neu.f90}, which is
used to simulate the effect of thermal neutrino emission. However, in this
approach, one has to be careful about the internal parameters \verb|log10Tmin_neu| and 
\verb|log10_Tlim|, which cause MESA to stop simulating extra
neutrinos when the temperature is below $10^{7}\ \text{K}$. As the core temperature of the stars considered here are below this threshold one needs to adjust
their values and re-compilation of the core MESA code to pursue this direction. If the
issue of \verb|log10Tmin_neu|  is handled properly, we have checked
that both approaches lead to the same results. In practice, we recommend
using  \verb|s% other_energy| instead of \verb|s% other_neu|. 

In order to accurately compute the stellar radius, we turn on the
atmosphere option: \nolinkurl{atm_option = 'table'} in \nolinkurl{inlist_project}. 

Our particle physics part of the calculation is conducted in natural
units while the MESA code uses cgs units. Hence, proper conversion
between the two unit systems is required. For instance, the luminosity obtained
in natural units is converted to cgs units via $1\text{eV}^{2}\approx 2434\ \text{erg}/\text{s}$. 

We start the evolution of a star from a pre-main-sequence model and
terminate it when the star reaches the typical age of a red dwarf,
which is set at $4.85\times10^{9}$ years in our code. Varying this
number from one billion (typical time for a red dwarf to enter the
main sequence) to 14 billion years (the age of the universe) causes
no noticeable effect on the final result.

\begin{table}[htb]
\centering
\caption{Data from Swayne et al \cite{2024MNRAS.528.5703S}.}
\label{tab:swayne}
\begin{tabular}{lcccccc}
\toprule
Target   & $M [M_\odot]$  & $\sigma_M$ & $R [R_\odot]$  & $\sigma_R$ & ${\rm Fe/H}$   & $\sigma_{\rm Fe/H}$ \\
\toprule
J0113+31 & 0.1974 & 0.0068     & 0.2193 & 0.0033     & $-0.31$ & 0.05       \\
J0540-17 & 0.1633 & 0.0058     & 0.1949 & 0.0032     & $-0.04$ & 0.05       \\
J0719+25 & 0.1584 & 0.0055     & 0.1917 & 0.0029     & 0.04    & 0.05       \\
J0941-31 & 0.2173 & 0.0078     & 0.2365 & 0.0036     & 0.078   & 0.069      \\
J0955-39 & 0.2211 & 0.008      & 0.2327 & 0.003      & $-0.24$ & 0.08       \\
J1305-31 & 0.282  & 0.0095     & 0.2982 & 0.0042     & 0.201   & 0.044      \\
J1522+42 & 0.1656 & 0.0063     & 0.1915 & 0.0043     & $-0.061$& 0.044      \\
J1928-38 & 0.2703 & 0.0091     & 0.2692 & 0.0057     & $-0.009$& 0.042      \\
J1934-42 & 0.196  & 0.0076     & 0.2241 & 0.0067     & 0.288   & 0.046      \\
J2040-41 & 0.1524 & 0.0053     & 0.1802 & 0.0032     & $-0.206$& 0.043      \\
J2046-40 & 0.1917 & 0.0067     & 0.2212 & 0.0046     & 0.337   & 0.054      \\
J2046+06 & 0.1769 & 0.0062     & 0.2055 & 0.0025     & 0.0     & 0.048      \\
J2315+23 & 0.2309 & 0.0099     & 0.2521 & 0.0034     & 0.02    & 0.05       \\
J2343+29 & 0.1202 & 0.0046     & 0.1464 & 0.0031     & 0.11    & 0.05       \\
J2359+44 & 0.293  & 0.010       & 0.2978 & 0.0036     & 0.12    & 0.05       \\
\toprule
\end{tabular}
\end{table}

\end{document}